\documentclass{eas}
\usepackage{graphicx}
\begin{document}

\title{Effects of Quasi-Static Aberrations in Faint Companion Searches} 
\runningtitle{Differential Simultaneous Imaging}
\author{Christian Marois, Ren\'{e} Doyon, Daniel Nadeau, Ren\'{e} Racine}\address{D\'{e}partement de physique, Universit\'{e} de Montr\'{e}al, C.P. 6128, Succ. A, Montr\'{e}al, QC, Canada H3C 3J7}
\author{Gordon~A.~H.~Walker}\address{1234 Hewlett Place, Victoria, BC, Canada V8S 4P7}
\begin{abstract}
We present the first results obtained at CFHT with the TRIDENT infrared camera, dedicated to the detection of faint companions close to bright nearby stars.  The camera's main feature is the acquisition of three simultaneous images in three wavelengths (simultaneous differential imaging) across the methane absorption bandhead at 1.6~$\mu $m, that enables a precise subtraction of the primary star PSF while keeping the companion signal. The main limitation is non-common path aberrations between the three optical paths that slightly decorrelate the PSFs. Two types of PSF calibrations are combined with the differential simultaneous imaging technique to further attenuate the PSF: reference star subtraction and instrument rotation to smooth aberrations. It is shown that a faint companion with a $\Delta H$ of 10 magnitudes would be detected at $0.5^{\prime \prime}$ from the primary.
\end{abstract}
\maketitle
\section{Introduction}
The next generation high order adaptive optics (NGAO) systems that are currently in development will produce high Strehl ratio images in the infrared on 10~m class telescopes. If those NGAO systems are to be used to search for very faint companions (brown dwarfs and exoplanets), it is of great importance to understand the limits for high contrast imaging on current adaptive optics (AO) systems and see how they would affect NGAO systems performances. Static aberrations have long been known to be the limiting factor for space-based telescopes when attempting high contrast imaging for faint companion detection (\cite{Brown90}). If the telescope physical conditions change in time (like point spread function (PSF) breathing for HST), the structure becomes quasi-static and produces PSF structures that evolve slowly in time, making difficult a good PSF calibration when using reference stars. Long AO corrected exposures have shown that ground-based imaging is not free from this problem. When using an AO system that delivers partially diffracted PSFs, the final wavefront is an interference between the AO wavefront residuals and the optical path wavefront aberrations to the detector. If enough random aberrations are removed by the AO system, a coherent PSF starts to appear. This is also true for the quasi-static portion of the wavefront. As more turbulence is removed, the quasi-static wavefront errors not corrected by the AO system will introduce coherent structures that will not average out in time and will mask faint companions. A partially-corrected long exposure AO observation thus suffers from the same problem that limits space telescopes. The current limit for detecting faint companions with existing AO systems is not the atmospheric turbulence correction efficiency but rather a quasi-static PSF calibration problem.
\vspace{0.3cm}

In the past few years, our group has developed a specialized infrared camera, TRIDENT (\cite{Marois2000a,Marois2002}), to overcome the PSF calibration problem. The main feature of TRIDENT is to acquire three simultaneous images in three distinct narrow spectral bands. It is possible to enhance the star/companion contrast after image combinations by selecting special spectral features that are typical of the companion and not of the star (\cite{Smith87,Rosen87,Racine99,Marois2000b}). In TRIDENT, the three wavelengths (1.567~$\mu$m, 1.625~$\mu$m and  1.680~$\mu$m, 1\% bandwidth) have been selected across the 1.6 $\mu$m methane absorption bandhead that is only present in the spectrum of cold ($T_{\rm{eff}}$ $<$ 1470~K, \cite{Feg96}) substellar objects. The 1.567~$\mu$m image ($I_{\lambda_1}$) shows the star and companion while the 1.625 and 1.680~$\mu$m images ($I_{\lambda_2}$ and $I_{\lambda_3}$) show the star with a fainter companion due to the methane absorption bandhead. The 1.625 and 1.680~$\mu$m images can thus be used as reference PSFs to correct the first and second order PSF structure evolution with wavelength. Since the three wavelengths are simultaneous, each observed wavelength sees the same PSF atmospheric distortion, a good PSF correlation between the three wavelengths can thus be achieved. This concept would remove atmospheric speckles and problems associated with the PSF evolution with telescope pointing, thermal changes and atmospheric r$_0$ variations. PSF subtraction is achieved by calculating simple and double image differences (SD and DD) obtained with the following algorithms (\cite{Marois2000b}):

\begin{center} $sd_{1-2} = I_{\lambda_1} - I_{\lambda_2}$ and $dd = sd_{1-2} - $k$*sd_{1-3}$
\end{center}

The DD can bring the atmospheric speckle noise below the level of the photon noise and remove the PSF quasi-static structure to its second order evolution with wavelength.
\clearpage
\section{Data acquisition and reduction}
Data were obtained on 2001 July 8-12 and on 2001 November 21-24, at the f/20 focus of the 3.6~m CFHT AO bonnette PUEO (\cite{Rigaut98}) with TRIDENT. Flat fields and darks were obtained during the day. In July, data well inside the linear regime of the detector were acquired to test the PSF stability. Total integration time was typically 1~h per target. Seeing conditions for this run were good (Strehl of 0.5 in the $H$ band). In November, reference stars were acquired and instrument rotations ($+/-$ 90 degrees in steps of 2 to 5 degrees) were done to calibrate or smooth common and non-common path aberrations discovered to be the limiting factor in the July mission data. Saturated and non-saturated images were acquired to minimize readout noise. Total integration time was 1h30 with medium seeing conditions (Strehl from 0.15 to 0.4 in the $H$ band). In total, 35 stars have been observed during these two missions with spectral type ranging from B to M. Some stars were observed on different nights to increase their signal to noise ratios and study PSF stability. More details on data reduction and analysis are given in Marois~et~al.~(2003a) and Marois~et~al.~(2003b).

\section{Simulations}
A long AO-corrected exposure is a superposition of a coherent PSF with a residual atmospheric halo. The coherent PSF does not look like a perfect diffraction pattern, but is distorted with some structures. Figure 1 shows the observed PUEO-TRIDENT PSF compared to a simulated PSF of 100 independent atmospheric speckle patterns (approximately 1.5~s exposure).

\begin{figure}[h]
\includegraphics[bb=-15 0 0 320,scale=0.5]{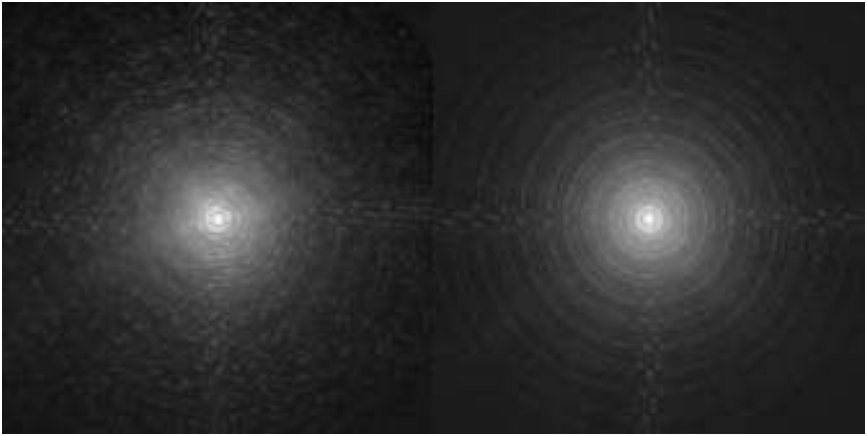}
\caption{Ups And (1~h integration, $6^{\prime \prime}$ field of view (FOV)) is shown in the left frame. The right frame shows a simulation of 100 independent speckle patterns (logarithm intensity scale). Photon noise and read noise are neglected.}
\end{figure}
\clearpage
The observed Ups And AO corrected PSF shows a series of broken rings with a diffuse atmospheric halo while the simulated PSF shows a series of perfect diffraction rings with the same diffuse halo. Structures seen in long exposure AO corrected PSFs are not due to turbulence but to a wide range of low to high frequency telescope and instrument aberrations. Increasing the integration time beyond a few seconds increases the signal to noise ratio of the coherent PSF structure. The total estimated quasi-static aberrations residual is 110~nm RMS or a $H$ band Strehl ratio of 0.82. This result is in good agreement with the measured value of the PUEO AO system (approximately 0.84 in the $H$ band, \cite{Rigaut98}), comfirming that most of the aberrations are coming from the PUEO AO system.  The estimated power spectrum has the following shape: a decreasing spectrum for the first 12 Zernike terms that correspond to 65\% of the total aberrations, a constant power spectrum for 150 Zernike terms that have 20\% of the total aberrations, a mid-frequency component that have 5\% of the total aberrations and a high frequency component with 10\% of the total aberrations. Figure 2 shows the same simulation as figure 1 (100 coadditions of independent atmospheric speckle patterns) but including, for each pattern, a static phase error that interferes with the atmospheric turbulence. The simulated PSF now shows some broken diffracted rings as observed in the Ups And PSF.

\begin{figure}[h]
\includegraphics[bb=-15 0 0 320,scale=0.5]{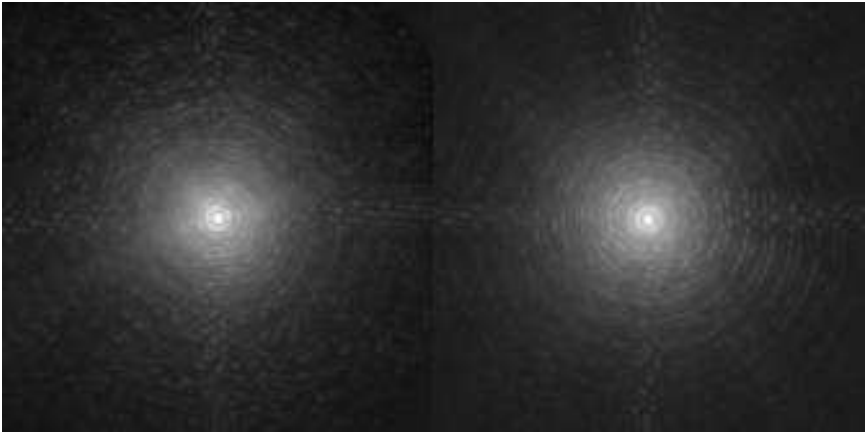}
\caption{Ups And (1~h integration, $6^{\prime \prime}$ FOV) is shown in the left frame. The right frame shows a simulation of 100 independent speckle patterns with a static phase error (logarithm intensity scale). Photon noise and read noise are neglected.}
\end{figure}

The simulations were done to determine the SD and DD subtraction performances. Images were magnified to the same number of pixels per Full Width Half Max to correct for the PSF chromatic dependence. The $3\sigma$ detection limits are presented in figure 4 and correspond to the theoretical limits of the TRIDENT differential simultaneous imaging technique with the current estimated static aberrations power spectrum.
\clearpage

\begin{figure}[h]
\includegraphics[bb=-120 0 0 370,scale=0.45]{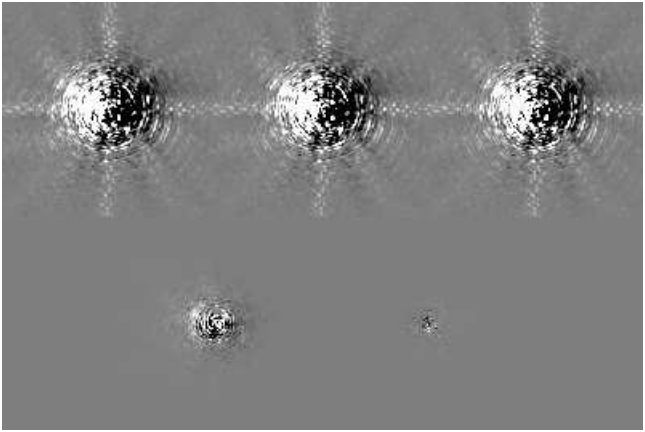}
\caption{Simulated PSF attenuation predictions. The first three upper frames are the PSFs at the three wavelengths (1.567, 1.625 and 1.680~$\mu $m) with their radial profiles removed (6$^{\prime \prime}$ FOV, $\pm 10^{-4}$ from PSF maximum and linear intensity scale). The lower two frames are (from left to right) the SD and DD with their radial profiles removed.}
\end{figure}

\begin{figure}[h]
\includegraphics[bb=-60 0 0 300,scale=0.63]{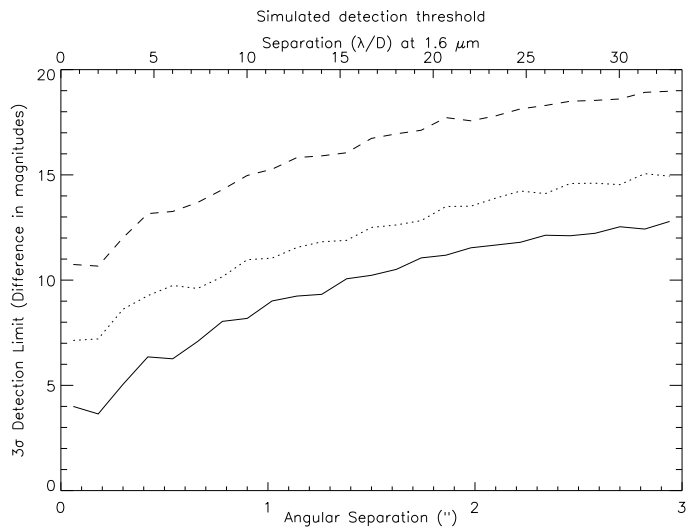}
\caption{$3\sigma$ detection limit vs angular separation for the simulated PSFs. The solid line represents the initial PSF minus its radial profile, the dotted line represents the SD and the dashed line represents the DD.}
\end{figure}
\clearpage

Simulated aberrations in figures 2 and 3 reproduce the power spectrum of the observed data. The asymetries that can be seen in figure~3 (three upper frames) come from the aberrations randomly generated in the simulation. A companion 10 magnitudes fainter than the primary at $0.5^{\prime \prime}$ separation would be detected with the SD alone, while a companion 14 magnitudes fainter than the primary would be detected at the same separation with the DD.

\section{Observations}
\noindent Figure 5 show the observed PSF attenuation SD and DD for Ups~And. Corresponding $3\sigma$ detection thresholds are presented in figure~6.

\begin{figure}[h]
\includegraphics[scale=0.705]{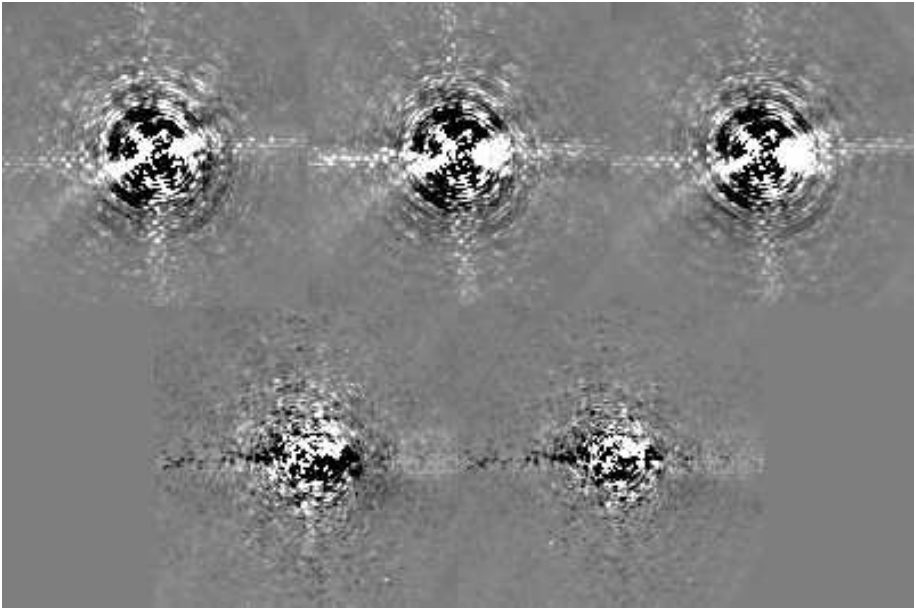}
\caption{Observed PSFs for November 2001 Ups And data set. The three upper frames are the PSFs at the three wavelengths with their radial profiles removed (6$^{\prime \prime}$ FOV, $\pm 10^{-4}$ from PSF maximum and linear intensity scale). The lower two frames are (from left to right) the SD and DD with their radial profiles removed.}
\end{figure}

\clearpage
\begin{figure}[h]
\includegraphics[scale=0.73]{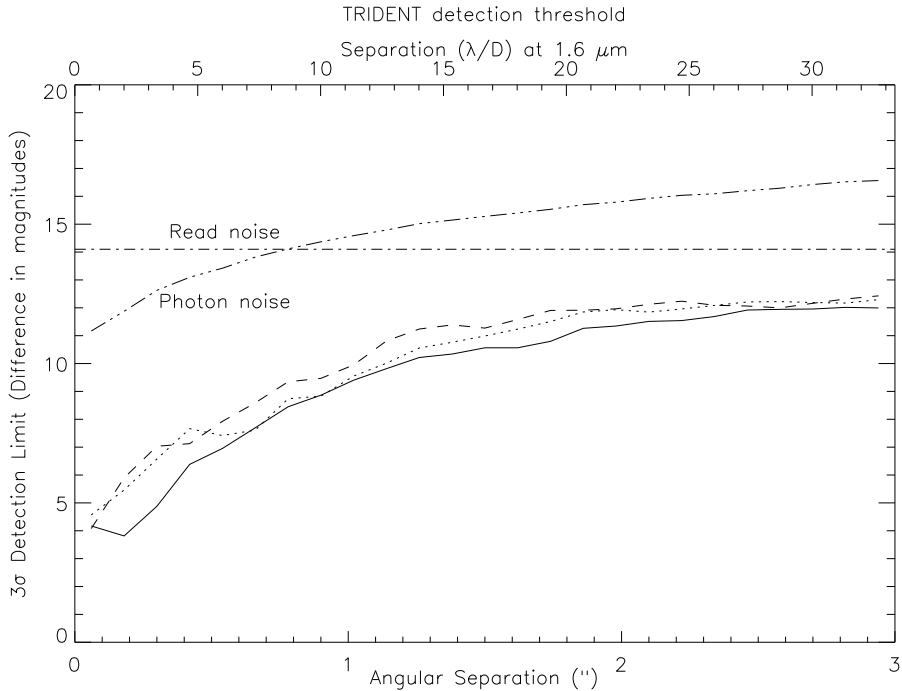}
\caption{Detection profile with angular separation for the November 2001 Ups And data set. The solid line represents the initial PSF with its radial profile removed, the dotted line represents the SD and the dashed line represents the DD.}
\end{figure}

\noindent As seen in figure 6, a companion 7 magnitudes fainter than the primary would be detected at $0.5^{\prime \prime}$ separation. The Ups~And SD and DD do not improve the detection limit contrary to simulated predictions (figure 4). Non-common path aberrations between the three optical paths (not included in the simulation presented in figure 4) are believed to interfere with common path aberrations (prior to TRIDENT beam splitter) and decorrelate the three PSFs. This shows the importance of minimizing non-common path aberrations when designing multiple optical path instruments.

\section{PSF calibration techniques}

PSF calibration techniques are needed to further reduce the PSF residual. A usual calibration technique consists of using other observations made during the same night but with different telescope orientations. This approach was found to be ineffective since the PSF structure varied slowly in time with telescope pointing and evolving atmospheric conditions. Figure 7 shows a set of July 2001 simple differences for a range of telescope pointings.
\clearpage
\begin{figure}[h]
\includegraphics[scale=0.58]{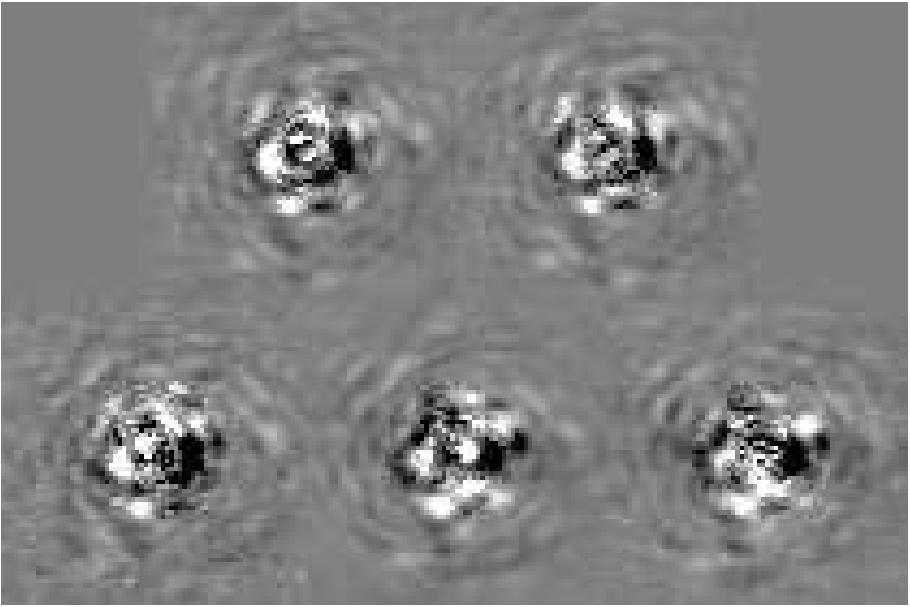}
\caption{PSF quasi-static structure evolution with changing telescope pointing. The two upper frames show the SD of the same object (Gl526) with its radial profile subtracted but on two different nights (July 8 and 9 2001) at the same local time. The lower three SD are respectively (from left to right) HIP111932 (July 8), HD187123 (July 8) and Gl614 (July 9) all acquired at different telescope pointing orientations (1.3$^{\prime \prime}$ FOV, $\pm 5\times 10^{-3}$ from PSF maximum and linear intensity scale).}
\end{figure}

\noindent The PSF is stable for a given telescope pointing (Gl526 data set from July 8 and 9 2001, see two upper frame in figure 7) but it is evolving when changing the telescope orientation (figure 7, three lower frames). This PSF evolution is probably generated by optical flexions with changing telescope orientation and the differential atmospheric refraction between the optical wavefront sensor and the infrared detector, making the three optical paths change with time. However, a much more accurate PSF calibration is obtained if a reference star is acquired at approximately the same telescope pointing orientation within a few minutes of the target star observations.

\subsection{Reference star subtraction technique}
To minimize the effect of PSF evolution with telescope pointing, the reference stars were selected at approximately the same celestial declination and 15 minutes East or West of the target star.
\clearpage
\begin{figure}[h]
\includegraphics[scale=0.73]{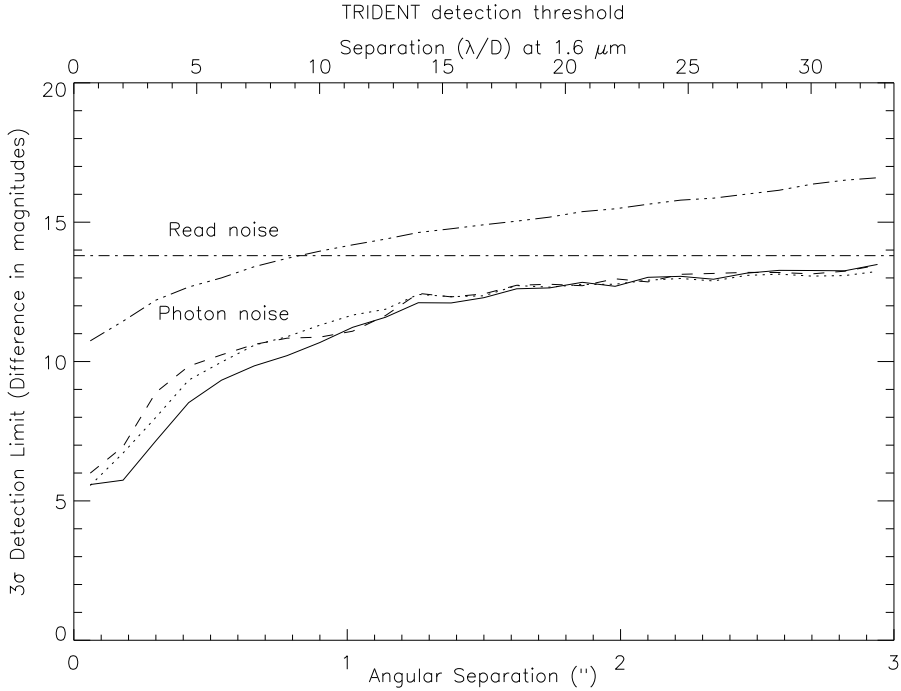}
\caption{Detection profile with angular separation for the November 2001 Ups And data set with the reference Chi And star subtracted. The solid line represents the initial PSF with its radial profile subtracted, the dotted line represents the SD and the dashed line represents the DD.}
\end{figure}

\noindent A typical gain of 3 magnitudes in detection sensitivity is obtained with the combination of reference star subtraction and simultaneous imaging (figure~8) as compared with simultaneous imaging alone (figure~6), thus confirming the quasi-static nature of the PSF structure since the reference and object were acquired within a few minutes. It can be seen that the PSF subtraction residuals are still 3 magnitudes away from being photon noise limited at 0.5$^{\prime \prime}$. Difficulty to calibrate optical aberrations due to PSF evolution with telescope pointing and atmospheric variations is the main limitation. A companion 10 magnitudes fainter than the primary would be detected at 0.5$^{\prime \prime}$ separation.

\subsection{Aberration smoothing by instrument rotation}
Another way to minimize PSF structure effects generated after the instrument rotator is to smooth them by rotating the instrument (PUEO AO system and TRIDENT camera). The PSF attenuation obtained by PSF subtraction after smoothing, with a Strehl ratio of 0.15, is similar to that obtained with reference star PSF subtraction for data having a Strehl ratio of 0.5, showing that aberrations were effectively smoothed by instrument rotation and thus confirming that they are coming mostly from PUEO and TRIDENT (Figure~9).

\begin{figure}[h]
\includegraphics[scale=0.73]{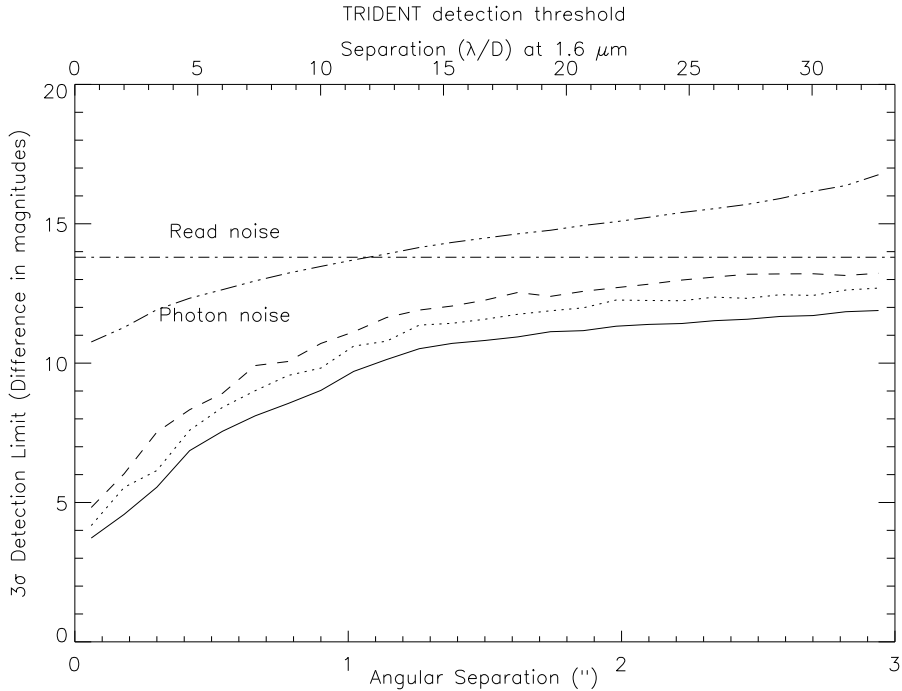}
\caption{Detection profiles with angular separation for the the November 2001 Ups And data with the instrument rotation technique. 35 rotations in steps of 5 degrees were combined using a median algorithm. The solid line represents the smoothed initial PSF with its radial profile subtracted, the dotted line represents the SD and the dashed line represents the DD.}
\end{figure}
\noindent The aberration smoothing technique has increased the correlation between the three optical paths by averaging non-common path aberration after the TRIDENT beam-splitter. The SD and DD with the PSF rotation technique (figure~9) are thus more efficient to subtract the PSF than differential imaging alone (figure~6). A companion 8.5 magnitudes fainter than the primary would be detected at 0.5$^{\prime \prime}$ separation.

Figure 10 shows the $\frac{3\sigma (I)}{I}$ PSF suppression efficiency (also known as the figure of merit, or Q factor) for the differential simultaneous imaging technique with reference star subtraction or aberration smoothing by instrument rotation.
\clearpage
\begin{figure}[h]
\includegraphics[scale=0.83]{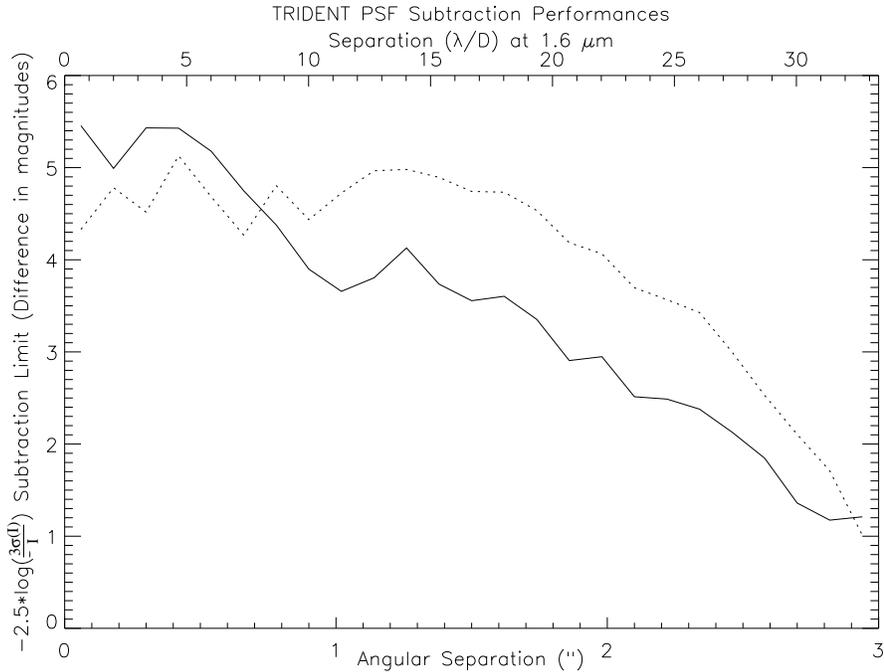}
\caption{$\frac{3\sigma (I)}{I}$ PSF suppression efficiency (or Q factor) in magnitude for the differential simultaneous imaging technique combined to the reference star subtraction (solid line) or the aberration smoothing technique by instrument rotation (dotted line).}
\end{figure}

\noindent Both the reference subtraction and aberration smoothing techniques deliver \linebreak[3] similar PSF subtraction limits. Reference subtraction is slightly better for separation smaller than 0.6$^{\prime \prime}$ while the aberration smoothing technique is slightly better for separations greater than 0.6$^{\prime \prime}$. The next step would be to combine both calibration techniques.

\section{TRIDENT new developments}
Photon noise limited subtraction requires the minimization of optical aberrations. To increase PSF correlation a new optical design is presently under evaluation that mainly consists of a better beam splitter and filters placed near the detector to minimize wavefront degradation and non-common path aberrations. The goal is photon noise limited performance for a few hours integrations.
\clearpage
\section{Conclusion - Quasi-static aberrations, an important factor}

Quasi-static aberrations in AO corrected images are the main problem when attempting high contrast imaging for faint companion searches. If quasi-static aberrations cannot be minimized or removed in NGAO systems, an optimized camera will be essential to calibrate and subtract them. If a multiple optical path camera is considered, non-common path aberrations need to be minimized to avoid PSF decorrelation. Reference star subtraction and aberration smoothing by instrument rotation improve by one order of magnitude the faint companion detection threshold. Typical attenuation in good seeing conditions are 10 magnitudes in $H$ band at $0.5^{\prime \prime}$ on a 3.6~m telescope.

\vspace{0.5cm}
This work is supported in part through grants from NSERC, Canada and from Fonds FQRNT, Qu\'{e}bec.

\end{document}